\documentclass[11pt,preprint]{aastex}
\usepackage{bm}
\usepackage{chngpage}
\usepackage{graphicx}
\usepackage[caption=false]{subfig}
\usepackage{natbib}
\usepackage{mathrsfs}

\def\apjl{ApJL}

\shorttitle{Anisotropic Jet}
\shortauthors{Geng et al.}

\begin{document}

\title{STEEP DECAY OF GRB X-RAY FLARES: THE RESULT OF ANISOTROPIC SYNCHROTRON RADIATION}

\author{Jin-Jun Geng\altaffilmark{1,2}, Yong-Feng Huang\altaffilmark{1,2}, Zi-Gao Dai\altaffilmark{1,2}}

\altaffiltext{1}{School of Astronomy and Space Science, Nanjing University, Nanjing 210023, China, hyf@nju.edu.cn, gengjinjun@gmail.com}
\altaffiltext{2}{Key Laboratory of Modern Astronomy and Astrophysics (Nanjing University), Ministry of Education, Nanjing 210023, China}

\begin{abstract}
When an emitting spherical shell with a constant Lorentz factor turns off emission abruptly at some radius,
its high-latitude emission would obey the relation of $\hat{\alpha}$ (the temporal index) = $2 + \hat{\beta}$ (the spectral index).
However, this relation is violated by the X-ray fares in some gamma-ray bursts (GRBs), whose $\hat{\alpha}$ is much more steeper.
We show that the synchrotron radiation should be anisotropic when the angular distribution of accelerated electrons
has preferable orientation, and this anisotropy would naturally lead to a steeper decay for the high-latitude emission
if the intrinsic emission is limb-brightened. We use this simple toy model to reproduce the temporal and spectral
evolution of X-ray flares. We show that our model can well interpret the steep decay of the X-ray flares in the three GRBs
selected as an example. Recent simulations on particle acceleration may support the specific anisotropic distribution of the electrons
adopted in our work. Reversely, confirmation of the anisotropy in the radiation would provide meaningful clues to the details of
electron acceleration in the emitting region.
\end{abstract}


\section{INTRODUCTION}

For relativistic astrophysical phenomena, such as gamma-ray bursts (GRBs), it is known that
the relativistic bulk motion would induce two significant effects on the radiation from their emission site when they are being observed.
First, due to relativistic boosting, the emission is beamed in the direction of motion.
So, for a jet with a certain opening angle (e.g., jets in GRBs or active galactic nuclei),
the emission from higher latitudes will have a smaller Doppler factor.
Second, due to the curvature of the geometry, photons at higher latitudes will arrive later than
those from the line of sight although they are emitted simultaneously \citep{Waxman97,Moderski00,Granot05,Huang07}.
The combination of these two effects is also called as the ``curvature effect'' for a relativistic spherical shell.
In other words, for a long-lasting, spherical emitting jet, the photons received by the observer at specific observer time actually
come from a distorted ellipsoid, rather than a spherical surface.

If the relativistic spherical shell flashes only at sometime, its temporal and spectral evolution of the light curve
has been predicted by previous researches.
Assuming the flux spectrum is a power-law form as $F_{\nu^{\prime}} \propto \nu^{\prime -\hat{\beta}}$ in the co-moving frame of the shell
and the bulk Lorentz factor of the shell $\Gamma$ is a constant, then the observed spectral flux would obey
$F_{\nu_{\rm obs}}^{\rm obs} \propto \nu_{\rm obs}^{-\hat{\beta}} t_{\rm obs}^{-\hat{\alpha}}$ and one can get the relation
$\hat{\alpha} = \hat{\beta} + 2$ together (e.g., \citealt{Kumar00,Dermer04,Liang06,Kumar15}),
where $\nu^{\prime}$ is the emitted frequency, $\nu_{\rm obs}$ is the observed frequency,
and $\hat{\alpha}$, $\hat{\beta}$ are the temporal index and spectral index respectively.
Hereafter, the superscript prime ($\prime$) is used to denote the quantities
in the co-moving frame and the letters ``$\rm{obs}$'' is used to denote the quantities in the observer frame.

After the GRB trigger, X-ray flares are often observed, thanks to the the X-Ray Telescope (XRT; \citealt{Burrows05}) on the {\it Swift} satellite \citep{Gehrels04}.
In general, X-ray flares show rapid rise and steep decay structures superposed on the underlying afterglow \citep{Zhang06}.
Several scenarios have been proposed for X-ray flares, including the clumpy accretion of the central engine \citep{Perna06},
the reconnection-driven explosive event from a post-merger neutron star \citep{Dai06}, the episodic accretion of the central black
hole \citep{Proga06}, or the delayed magnetic dissipation of the outflow \citep{Giannios06}, etc (see \citealt{Kumar15} for a review).
Although the proper model for X-ray flares is still uncertain, some studies suggest that X-ray flares and the gamma-ray prompt emission
may share a common origin, which further indicates that X-ray flares come from relativistic jets \citep[e.g.][]{Chincarini07,Lazzati07,Maxham09,Margutti10}.
The X-ray flares may be released either by the dissipation of
the magnetic energy \citep{Meszaros97,Zhang11}, or the kinetic energy of the jet \citep[e.g.][]{Paczynski86,Shemi90,Rees92}.
On the other hand, the decay phase of the X-ray flare should be the consequence of the cease of the energy release at the emitting site.
Therefore, X-ray flares may be a well benchmark to test the curvature effect for a relativistic spherical shell.

Indeed, \cite{Uhm15} have shown that the relation $\hat{\alpha} = \hat{\beta} + 2$ is invalid in the steep
decay phase of some X-ray flares. Furthermore, they pointed out that this invalidation may be the evidence that the
X-ray flare emission region is undergoing rapid bulk acceleration \citep{Uhm16a,Jia16}.
However, except for the bulk acceleration, there is another potential effect--the anisotropy of the radiation\footnote{
In this paper, by saying the anisotropy of the radiation, we mean that the emissivity of the
emitting electrons would has anisotropic angular distribution averagely in the co-moving frame, rather than
the anisotropic characteristics of jet's properties (e.g., \citealt{Dai01}).},
that can change the standard relation.
A main assumption above is that the radiation in the co-moving frame is isotropic, which still remains unconfirmed.
If the radiation in the co-moving frame is anisotropic, i.e., the radiation is latitude dependent,
then the decay phase would be determined by both the curvature effect and the intrinsic anisotropic characteristics.
In fact, the steep decay induced by the anisotropy have been revealed in the afterglow \citep{Beloborodov11}
and the prompt emission \citep{Barniol16,Beniamini16a,Granot16}.
One possible origin for the anisotropy is the preferable relative orientation between the direction of accelerated
electrons and the magnetic field.
Therefore, it is crucial to see whether the anisotropy can interpret the steep decay of the X-ray flares.

In our study, we select X-ray flares in three GRBs as example, of which the flare structures are clear and the data are
of high quality. We show that the steep decay can be well explained by considering the anisotropy in the radiation.
Our paper is organized as follows. In Section 2, we present the analytical derivation for the observed spectral
flux in the case of anisotropic synchrotron emission.
In Section 3, we develop a simple model to do numerical calculation and show how the anisotropy would
reproduce the temporal and spectral evolution of these X-ray flares.
The conclusions are summarized in Section 4.

\section{RADIATION FROM A THIN SHELL}
Like GRB's prompt emission, X-ray flares are expected to be produced when the jet's
magnetic or kinetic energy is released. The emission region is far from the GRB central engine and
can be treated as a part (limited opening angle) of an expanding spherical shell.
Here, we take the synchrotron radiation as the main emission mechanism in X-ray flares
and analytically derive the light curve from the shell.
The spectral emissivity of a single electron of Lorentz factor $\gamma_e$ at frequency $\nu^{\prime}$
in the fluid rest frame is given by \citep{Rybicki79}
\begin{equation}
P_{\nu^{\prime}}^{\prime} = \frac{\sqrt{3} q_e^3 B^{\prime} \sin \alpha}{m_e c^2} F \left( \frac{\nu^{\prime}}{\nu_c^{\prime} \sin \alpha} \right)
=P_0^{\prime} \sin \alpha F \left( \frac{\nu^{\prime}}{\nu_c^{\prime} \sin \alpha} \right),
\end{equation}
where $q_e$ is electron charge, $m_e$ is electron mass, $c$ is the speed of light,
$\alpha$ is the pitch angle between the direction of the electron's velocity and the local rest frame
magnetic field $\bm B^{\prime}$, $F$ is the synchrotron spectrum function\footnote{The synchrotron spectrum function is defined as $F(x) = x \int_{x}^{+ \infty} K_{5/3}(k) d k$, where $K_{5/3}(k)$ is the Bessel function.}, and $\nu_c^{\prime} = 3 q_e B^{\prime} \gamma_e^2 / (4 \pi m_e c)$.
One can easily note that the term $\sin \alpha$ in Equation (1) indicates the dependence of the emissivity
on $\alpha$, which would further introduce the anisotropy in the radiation of a group of electrons as shown below.

For relativistic jets launched from the central rotating compact objects, the magnetic fields in them
are expected to be mainly toroidal beyond the light cylinder \citep{Lyubarsky09,Bromberg16}.
Also, the radial expansion of the jet would suppress the longitudinal component of the magnetic fields.
So, in our modelling, we assume the magnetic fields $\bm{B^{\prime}}$ are transverse to the jet's direction and they are tangled
in the local shock plane.
On the other hand, the electron distribution may be anisotropic, i.e., the angular distribution of electron moving directions is assumed to
be described by a function $f(\alpha)$, which gives
\begin{equation}
\frac{{\rm d} N_e}{\rm{d} \Omega_e^{\prime}} = N_{\rm tot} \frac{f(\alpha)}{4 \pi},
\end{equation}
where $N_{\rm tot}$ is the total number of the electrons and $f(\alpha)$ is normalized by $\int f(\alpha) \rm{d} \Omega_e^{\prime} = 4 \pi$.
By averaging $P_{\nu^{\prime}}^{\prime}$ on random $B^{\prime}$ in the shock plane, we can obtain the effective
spectral emissivity per solid angle for a single electron as
\begin{equation}
\frac{{\rm d} \bar{P_{\nu^{\prime}}^{\prime}}}{{\rm d} \Omega^{\prime}} \simeq  \frac{A(\theta^{\prime})}{4 \pi} P_0^{\prime} F \left( \frac{\nu^{\prime}}{\nu_c^{\prime} A(\theta^{\prime})} \right),
\end{equation}
with
\begin{equation}
A(\theta^{\prime}) =
\frac{1}{2 \pi} \int_{0}^{2 \pi} \left( 1 - \sin^2 \theta^{\prime} \cos^2 \phi \right)^{1/2} f \left[ \arccos \left( \sin \theta^{\prime} \cos \phi\right) \right] {\rm d} \phi,
\end{equation}
where $\theta^{\prime}$ is the angle between the direction of emitted photons and the local radial direction
and the geometric relation $\cos \alpha = \sin \theta^{\prime} \cos \phi$ has been used.
For approximation, the term $A(\theta^{\prime})$ emerges in the coefficient and the spectral function $F$ separately
in Equation (3).

For a group of electrons which obey a spectrum of $d N_e /d \gamma_e$, the total spectral power from them should be
\begin{equation}
\frac{{\rm d} P_{\nu^{\prime},{\rm tot}}^{\prime} }{{\rm d} \Omega^{\prime}} = \int \frac{{\rm d} \bar{P_{\nu^{\prime}}^{\prime}}}{{\rm d} \Omega}
\frac{{\rm d} N_e}{{\rm d} \gamma_e} {\rm d} \gamma_e \approx N_{\rm tot} \frac{A(\theta^{\prime})}{4 \pi} P_0^{\prime} G\left( \frac{\nu^{\prime}}{\nu_c^{\prime} A(\theta^{\prime})} \right).
\end{equation}
In the last equality of Equation (5), we introduce the function $G(x)$ to approximate the integral of electron
spectrum and to simplify the calculation. In practice, $G(x)$ should have a prior form according to the observations,
such as a ``Band-function'' shape \citep{Band93}.

\subsection{Light Curve}
We assume electrons in a spherical shell of radius $r$ instantaneously emit photons in a very short time interval $\delta t$
measured in the burst frame ($\delta t \ll \delta t_{\rm ring}$, $\delta t_{\rm ring}$ is defined below).
The shell is expanding with a bulk Lorentz factor $\Gamma$. An observer will first see photons emitted along the line of sight with $\theta = 0$
($\theta$ denotes the latitude of the region on the shell).
If we set the observer time $t_{\rm obs}$ equal to zero when receiving the photons emitted from $\theta = 0$, then
photons from a location of $\theta$ will be detected by the observer at observer time
\begin{equation}
t_{\rm obs} = \frac{r}{c} (1-\mu) (1+z),
\end{equation}
where $\mu = \cos \theta$, and $z$ is the redshift of the burst.

The number of electrons in the ring of $[\theta,\theta+\delta \theta]$ is $\delta \mu N_{\rm tot} / 2$
($N_{\rm tot}$ is the total number of electrons of the shell).
In the local burst frame, the specific spectral energy $\delta E_{\nu}$ emitted into the solid angle $\delta \Omega$ in $\delta t$
can be related to the quantities in the co-moving frame by
\begin{equation}
\frac{\delta E_{\nu}}{\delta t \delta \Omega} = \frac{\mathcal{D}^2}{\Gamma} \frac{\delta E_{\nu^{\prime}}^{\prime}}{\delta t^{\prime} \delta \Omega^{\prime}} = \frac{\mathcal{D}^2}{\Gamma} \frac{{\rm d} P_{\nu^{\prime},{\rm tot}}^{\prime} }{{\rm d} \Omega^{\prime}},
\end{equation}
where $\mathcal{D} = \Gamma^{-1} (1-\beta \mu)^{-1}$ is the Doppler factor.
When the observer sees the ring, the corresponding energy is $\delta E_{\nu,{\rm ring}} = \delta E_{\nu} \delta \mu / 2$,
the corresponding time duration is $\delta t_{\rm ring} = r \delta \mu / c$.
The spectral luminosity is thus $\delta L_{\nu} = \delta E_{\nu,{\rm ring}} / \delta t_{\rm ring}$.
For an observer at distance $D_L$, the observed flux is $\delta F_{\nu_{\rm obs}}^{\rm obs} = \frac{(1+z) \delta L_{\nu}}{D_L^2 \delta \Omega}$,
which can be further expressed as (also see \citealt{Uhm15})
\begin{equation}
\delta F_{\nu_{\rm obs}}^{\rm obs} = \frac{1+z}{4 \pi D_L^2} \frac{c}{2 r} \delta t \frac{A(\theta^{\prime}) N_{\rm tot} P_0^{\prime} G((1+z) \mathcal{D}^{-1}
A(\theta^{\prime})^{-1} \nu_{\rm obs} / \nu_c^{\prime})}{\Gamma^3 (1-\beta \mu)^2}
\end{equation}
by combining Equations (5-7). Note that Equation (8) is only valid for the shell which flashes once (within $\delta t$),
while it is analytically useful since we only focus on the steep decay phase here.
When calculating the observed flux from a continually emitting shell, one should integrate
the differential flux over the equal-arrival-time surface \citep{Waxman97,Granot99,Huang00}, or equally $\delta t$ in Equation (8),
which will be done in Section 3.

For $G(x) \propto x^{- \hat{\beta}}$, we can obtain
\begin{equation}
\delta F_{\nu_{\rm obs}}^{\rm obs} \propto A(\theta^{\prime})^{1+\hat{\beta}} t_{\rm obs}^{-(2+\hat{\beta})} \nu_{\rm obs}^{-\hat{\beta}},
\end{equation}
where $A(\theta^{\prime}) = A(\theta^{\prime}(t_{\rm obs}))$ can be calculated by considering the relation
$\cos \theta^{\prime} = (\cos \theta - \beta)/(1 - \beta \cos \theta)$ and Equation (6).
One can find that the flux is affected by the factor $A(\theta^{\prime}(t_{\rm obs}))$,
which is further determined by the function $f$.

\subsection{Anisotropic Case}
In isotropic case, $A(\theta^{\prime}) = 1$, Equation (9) can recover the relation $\hat{\alpha} = 2 + \hat{\beta}$.
However, for anisotropic case, the relation will not hold.
Here, for instance, we consider the limb-brightened case, i.e., the emission is strong at $\theta^{\prime} = \pi/2$
and weak near $\theta^{\prime} = 0$.
For photons emitted at $\theta^{\prime} = \pi/2$, the angle between these photons and the radial direction
in the observer frame is $\sim 1/\Gamma$. Therefore, most of the emission comes from a ring of angle
$\theta = \Gamma^{-1}$ and the peak will be delayed comparing with the isotropic source \citep{Barniol16}.

On the other hand, if we assume $A(\theta^{\prime}) \propto (\sin \theta^{\prime})^n$, $n>0$, then
using $\cos \theta^{\prime} \simeq (\tau - t_{\rm obs})/(\tau + t_{\rm obs})$, we have
\begin{equation}
\delta F_{\nu_{\rm obs}}^{\rm obs} \propto \left[ \frac{4 \tau t_{\rm obs}}{(\tau + t_{\rm obs})^2} \right]^{(1+\hat{\beta}) n / 2} t_{\rm obs}^{-(2+\hat{\beta})},
\end{equation}
where $\tau = r (1+z) / (2 \Gamma^2 c)$, and $\beta \simeq 1 - \frac{1}{2 \Gamma^2}$ is used.
We will have the temporal decay index as
\begin{equation}
\lambda = - \frac{d \ln F_{\nu_{\rm obs}}^{\rm obs}}{d \ln t_{\rm obs}} \simeq (1+\hat{\beta}) n / 2 + (2+\hat{\beta}),
\end{equation}
where $t_{\rm obs} \gg \tau$ is used.
According to Equation (11), the anisotropy of the radiation will lead to a steep decay.

\section{APPLICATION TO X-RAY FLARES}
Based on the analysis above, we now propose a simple model and perform numerical calculations
to reproduce the temporal and spectral behaviors of the X-ray flares.
In order to achieve the steep decay, we choose a function $f$ which corresponds to a limb-brightened case,
i.e., $f(\alpha) \propto (a^2 + \sin^2 \alpha)^{-3}$,
where $a$ is the characteristic beaming angle of the electron distribution. This expression
can be achieved when the electrons are preferentially moving along $\bm B$ and the resulting
expression for $A(\theta^{\prime})$ is limb-brightened.

To perform the calculations, we need to notice the ``timing'' of the data of X-ray flares.
Looking at the X-ray light curve, one may find one point ($t_{\rm obs} = T_0$)
after which the rising phase of the flare emerges.
If we reset the reference time ($t_{\rm obs} = 0$) at time $T_0$, then the burst-frame time $t$ can be
connected to the observer-frame time $t_{\rm obs}$ by
\begin{equation}
t_{\rm obs} = \frac{1}{c} [r_{\rm s} + c (t - t_{\rm s}) -r \cos \theta] (1 + z),
\end{equation}
where the initial photons of the flare are emitted at $r_{\rm s}$ at $t_{\rm s}$,
and $t = t_{\rm s} + \int_{r_{\rm s}} d r/ (c \beta)$.
However, the true value of $T_0$ may be obscured by the prompt emission
and its precision is limited by the timing resolution of the observation.
In practice, we use another parameter $\Delta T$ to describe the missed portion as is done in \cite{Uhm15}, so that the
observer time of the flare is
\begin{equation}
t_{\rm obs} = \frac{1}{c} [r_{\rm s} + c (t - t_{\rm s}) -r \cos \theta] (1 + z) - \Delta T.
\end{equation}

Moreover, the shape of the spectrum $G(x)$ should be given since we do not focus on detailed
radiation mechanism in this paper.
The spectrum of prompt emission is usually describe as a ``Band'' function \citep{Band93}.
However, the rapid softening of the spectrum during the decay phase of the X-ray flare indicates
that the spectrum may be a power-law with an exponential cutoff (also see \citealt{Uhm16a}).
We thus take $G(x) = x^{\zeta+1} e^{-x}$ in the following calculations.

In the co-moving frame, the total number of radiating electrons in the shell is $N_{\rm shell} = 0$ at the starting radius $r_{\rm s}$,
and is assumed to increase at a rate $R_{\rm inj}$ before the turn-off radius $r_{\rm off}$.
In our calculations, we model the evolution of the characteristic Lorentz factor $\gamma_{\rm ch}$,
and $R_{\rm inj}$ as
\begin{eqnarray}
\gamma_{\rm ch} (r) = \gamma_{\rm ch}^0 \left( \frac{r}{r_{\rm s}} \right)^g, \\
R_{\rm inj} (r) = R_{\rm inj}^0 \left( \frac{r}{r_{\rm s}} \right)^{\eta},
\end{eqnarray}
where $\gamma_{\rm ch}^0$ and $R_{\rm inj}^0$ are the initial value of $\gamma_{\rm ch}$ and $R_{\rm inj}$ at $r_{\rm s}$.
The indices $g$, $\eta$ describes how the characteristic Lorentz factor and the injection rate evolve with radius $r$
respectively. They are essential to model the rapid rise of the X-ray.
In addition, $r_{\rm s} = 10^{14}$ cm is commonly adopted in all our calculations.
We then integrate the flux from a series of rings of which the emitted photons reach the observer at the same time
to obtain the light curve of the X-ray flare.
In our calculations, the redshifts of selected GRBs are assumed to be $z = 1$
and the standard $\Lambda$CDM universe with $H_0 = 71$ km~s$^{-1}$~Mpc$^{-1}$, $\Omega_{\rm m} = 0.27$,
and $\Omega_{\Lambda} = 0.73$ is adopted.

Three GRBs are selected as examples, i.e., GRB 090621A, GRB 121108A, and GRB 140108A,
of which the X-ray data show significant rapid rise and steep decay.
The numerical results are shown in Figure 1, in comparison with the observed
light curve and photon index $\hat{\Gamma}$ ($\hat{\Gamma} = \hat{\beta}+1$).
The corresponding parameters in each case are listed in Table 1.
Our results from numerical calculations are in good agreement with the observations.
Note that the numerical light curves become steeper than the observed ones at the late stage of the decay phase.
However, this deviation does not change our main conclusion. The deviation can be understood in two aspects.
There is an obvious turning point in the evolution curve of the observed photon index at the late stage of the decay phase,
which strongly indicates that another flare component (or some intrinsic variabilities) should emerge and dominate at
that moment.
We are only modelling one component of the X-ray light curve, while the observational data do not come purely from
one single component.
On the other hand, our model is a toy model with some simplified relations, such as Equations (14-15),
which makes us unable to model additional variabilities shown in the observational data.
If we calculate the flux for the other component and add it to the result, the new total flux
would fit the data better.

\begin{deluxetable}{cccc}
\tabletypesize{\scriptsize}
\tablewidth{0pt}
\tablecaption{Parameters used in the modelling of the X-ray flares of the three GRBs.\label{TABLE:Modelling}}

\tablehead{%
        \colhead{Parameters} &
        \colhead{GRB 090621A} &
        \colhead{GRB 121108A} &
        \colhead{GRB 140108A}
        }
\startdata
$a$                                                           &    0.15      &      0.08        & 0.11 \\
$\zeta$                                                     &    -0.58     &      -0.35       & -0.48  \\
$B_0$$^{\dagger}$ (G)                            &    100.0     &     100.0      & 100.0  \\
$\Gamma$                                               &    30.0       &       23.0      & 22.0  \\
$\gamma_{\rm ch}^0$ ($10^4$)              &    2.4         &      0.6          & 1.2  \\
$g$                                                          &    0.0         &      1.35        & 0.7  \\
$\eta$                                                      &    0.0         &      0.7          & 0.0  \\
$\Delta T$ (s)                                          &    10.0       &      2.0          & 6.5  \\
$R_{\rm inj}^0$ ($10^{48}$ s$^{-1}$)      &    3.2         &      3.6          & 22.0  \\
$r_{\rm off}$ ($10^{14}$ cm)                   &    10.0       &      3.65          & 3.3
\enddata
\tablenotetext{\dagger}{$B_0$ is the strength of the rest frame magnetic field $\bm B^{\prime}$.}
\end{deluxetable}

\section{DISCUSSION}
In this study, we consider the anisotropy of the synchrotron radiation in the high-latitude emission and apply it
to the observed X-ray flares in three GRBs, i.e., GRB 090621A, GRB 121108A, and GRB 140108A.
The steep decay phase can be well interpreted by our scenario in which the intrinsic radiation
is limb-brightened. Also, the entire temporal, spectral behavior have been modelled using our
simplified model.
This is the first evidence that intrinsic synchrotron radiation from the emission site
of the X-ray flare may be anisotropic.

The anisotropy of the radiation in the co-moving frame is physically supported.
In general, the relativistic reconnection sites are expected to be the main sources of non-thermal electrons
in the Poynting-flux-dominated flow.
Recent particle-in-cell simulations on particle acceleration in the relativistic reconnection current sheet
have revealed that the electrons would be efficiently accelerated by the motional electric field when
they bounce back and forth within a magnetic island (see \citealt{Guo14,Guo15} for detailes).
The acceleration along the motional electric field (perpendicular to the X-line plane of the reconnection)
would naturally give a preferable angular distribution for electrons.
This preferable angular distribution is consistent with the function $f(\alpha)$
used in our work.
Thus the identification of the role the anisotropy plays in the steep decay of X-ray flares would give useful
clues for the details of particle acceleration in the emitting region.

It has been proposed that the steep decay of the X-ray flares may be the evidence for the emission
site being accelerating. However, in our work, we attribute the steep decay to the anisotropy of the radiation,
rather than the acceleration of the jet. Both mechanisms can explain some current observational features.
Observations on the polarisation of X-ray flares may help to identify the prior model
since the preferable relative orientation between the $\bm B^{\prime}$ field and the electrons moving direction in our model
may lead to a polarization degree different from the other model (being prepared).
On the other hand, since both mechanisms can coexist naturally in the same frame work (the jet is Poynting-flux dominated),
the possibility that they work together within the sample selected can not be neglected.
In \cite{Uhm16b}, they found that the acceleration of the emission region is needed to interpret spectral lags in the prompt emission,
but a shallower acceleration index is required, which suggests that the X-ray flare decay may have a contribution
from the anisotropic effect.
\cite{Beniamini16b} suggests that the material producing the X-ray flare may be confined to a jet
which is narrow compared to $1/\Gamma$, this provides another possible solution to the steep decay.

An anisotropic minijets model has been invoked to explain the short time-scale variability of the GRB prompt
emission \citep{Barniol16}, in which the radiation is also anisotropic in the co-moving frame.
Relevant works show that minijets are essential in defining light curves of the prompt emission \citep{Zhang14,Deng15,Deng17}.
The anisotropic characteristics in the GRB emission mechanism thus seem to be common
and need more researches.

\acknowledgments
We thank the referee, Prof. Pawan Kumar for very valuable suggestions.
We also thank Liang Li for helpful discussion.
This work is partially supported by the National Basic Research Program (``973'' Program) of China
(grant No. 2014CB845800) and the National Natural Science Foundation of China (grants No. 11473012 and 11573014).
The authors acknowledge support by the Strategic Priority Research Program of the Chinese Academy of Sciences
(``Multi-waveband Gravitational Wave Universe'' with Grant No. XDB23040000).
This work made use of data supplied by the UK Swift Science Data Center at the University of Leicester.

\begin{figure}
\centering
    \subfloat{\includegraphics[width=0.48\linewidth]{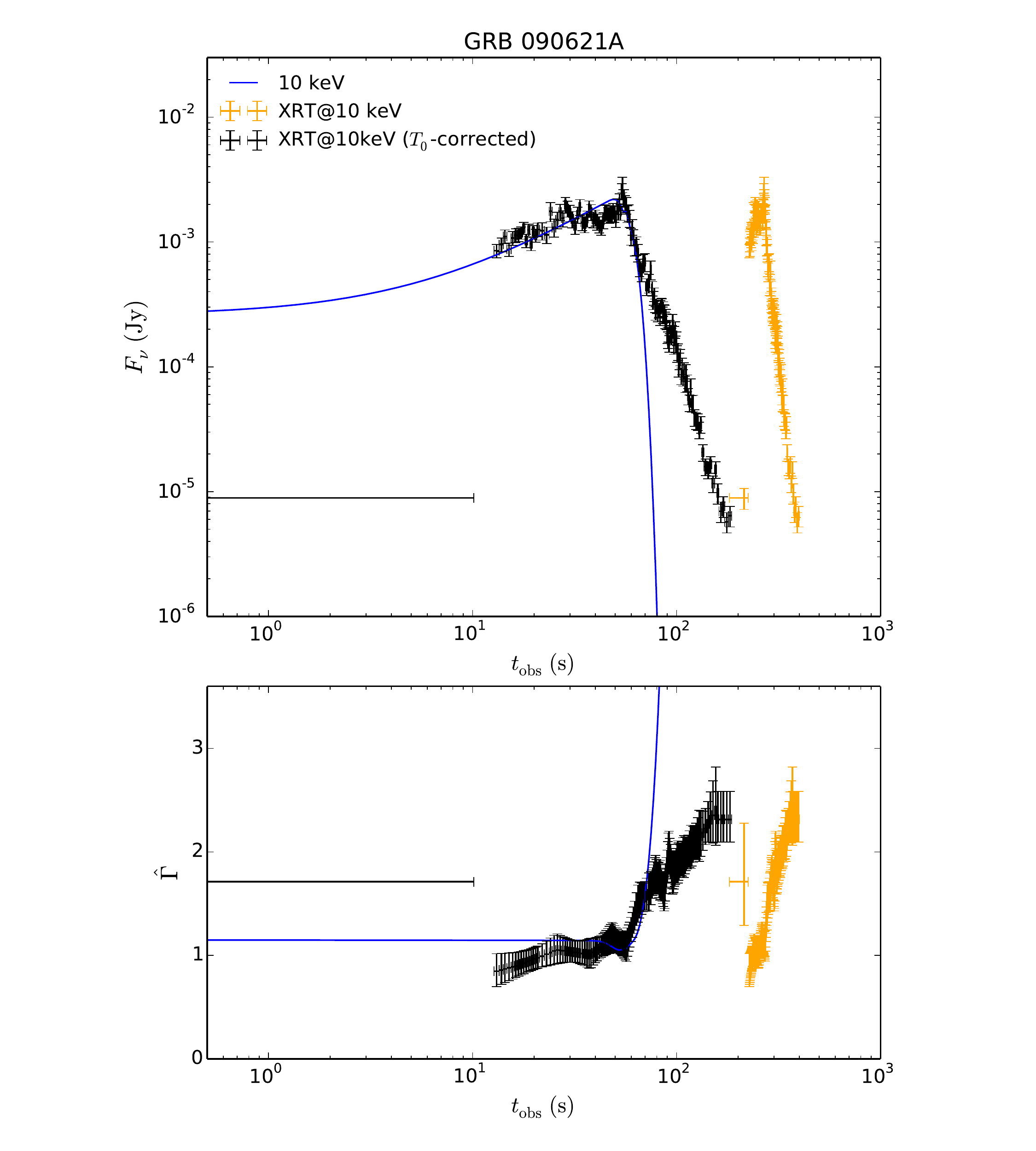}}
    \subfloat{\includegraphics[width=0.48\linewidth]{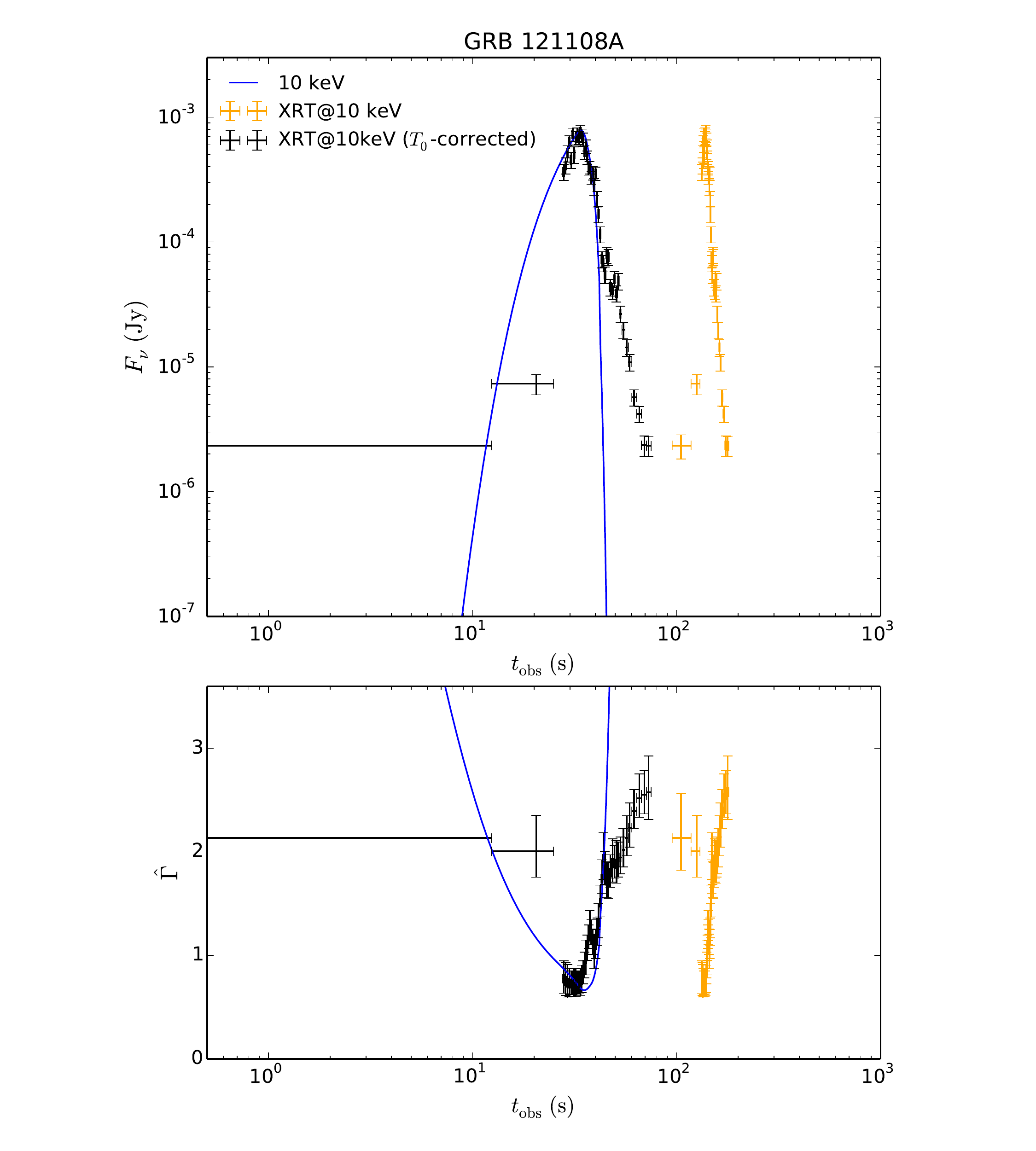}} \\
    \subfloat{\includegraphics[width=0.48\linewidth]{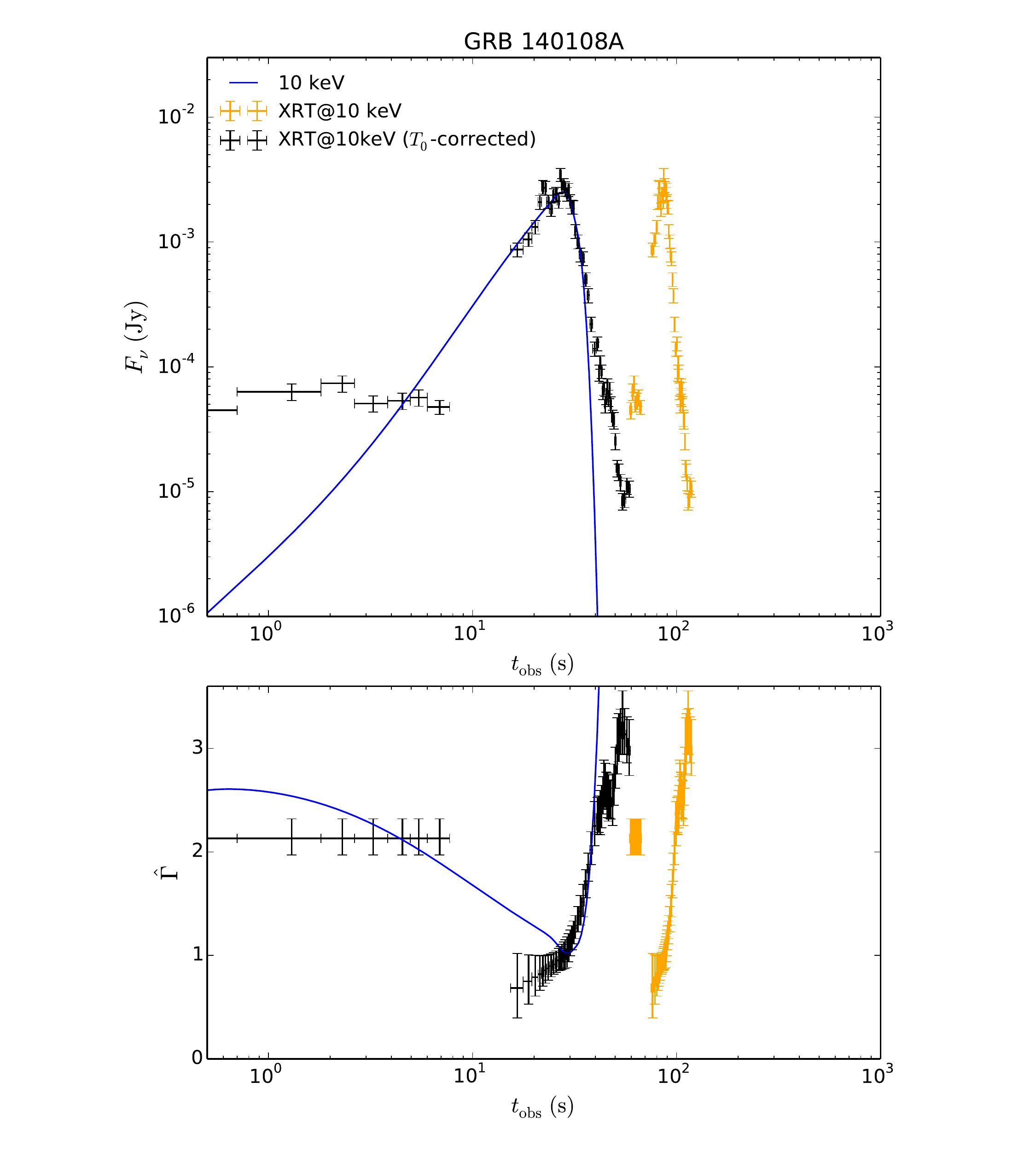}}
\caption{Modeling the X-ray flares in GRB 090621A, GRB 121108A and GRB 140108A by using the anisotropic radiation scenario.
For each GRB, in the upper panel, the original observed light curve at 10 keV (the orange points) is shown.
The black points are the ``shifting'' version\protect\footnotemark~of the original data by setting the reference time at $T_0$ (see details in Section 3)
and the model-calculated light curve is presented as a blue line.
The lower panel is similar to the upper panel, but presents the corresponding XRT band (0.3--10 keV) photon index.
}
\label{Fig:plot1}
\end{figure}
\footnotetext{This approach was presented in \cite{Uhm16a} to model the data for the first time.}

\end{document}